\documentclass{mem}
\usepackage{natbib}\usepackage{txfonts}\usepackage{balance}
\usepackage{graphicx}
\usepackage{subfigure}
\usepackage[a4paper]{hyperref}
\idline{75}{282}
\usepackage[english]{babel}

\def \xmmn {\emph{XMM-Newton }}

\def \chandra {\emph{Chandra }}

\begin{document}
 \title{A radio approach to the cool core -- non cool core dichotomy}

   \subtitle{}
	\author{M. \,Rossetti\inst{1}
\and B.\,M.\, Cavalleri\inst{2,1}
\and S.\, Molendi\inst{1}
\and F.\,Gastaldello\inst{1}
\and S.\,Ghizzardi\inst{1}
\and D.\,Eckert\inst{1}
}
\institute{
IASF-Mi -- Istituto Nazionale di Astrofisica
via Bassini 15, Milano, Italy
\email{rossetti@iasf-milano.inaf.it}
\and
Dipartimento di Fisica -- Universit\`a degli Studi di Milano
via Celoria 16, Milano, Italy
}
\authorrunning{Rossetti et al}
\titlerunning{A radio approach to the CC--NCC dichotomy}
\abstract{From the point of view of X-ray astronomers, galaxy clusters are usually divided
 into two classes: ``cool core'' (CC) and ``non-cool core'' (NCC) objects.
The origin of this dichotomy has been subject of debate in recent years, between
 ``evolutionary'' models (where clusters can evolve from CC to NCC, mainly through
 mergers) and ``primordial'' models (where the state of the cluster is fixed ``ab initio'' by early mergers or pre-heating).
We found that in a representative sample (clusters in the GMRT Radio halo survey with available X-ray data), none of the objects hosting a giant radio halo can be classified as a cool core.
This result suggests that the main mechanisms which can produce the ingredients
to start a large scale synchrotron emission (most likely mergers) are the same that can destroy CC  and therefore strongly supports ``evolutionary'' models of the CC-NCC dichotomy.
}
\maketitle{}

\section{Introduction}
Galaxy clusters are often divided by X-ray astronomers into two classes: ``cool core''(CC) and ``non-cool core'' (NCC) objects on the basis of the observational properties of their central regions. 
One of the open questions in the study of galaxy clusters concerns the origin of this distribution. 
The original model which prevailed for a long time assumed that the CC state was a sort of ``natural state'' for the clusters, and the observational features were explained with the old ``cooling flow'' model: radiation losses cause the gas in the centers of these clusters to cool and to flow inward. Clusters were supposed to live in this state until disturbed by a ``merger''. Indeed, mergers are very energetic events that can shock-heat \citep{burns97} and mix the ICM \citep{gomez02}: through these processes they were supposed to efficiently destroy cooling flows. After the mergers, clusters were supposed to relax and go back to the cooling flow state in a sort of cyclical evolution (e.\,g.\, \citealt{buote02}).
With the fall of the ``cooling flow'' brought about by the \xmmn and \chandra observations (e.\,g.\, \citealt{peterson01}), doubts were cast also on the interpretation of mergers as the dominant mechanism which could transform CC clusters into NCC. These doubts were also motivated by the difficulties of numerical simulations in destroying simulated cool cores with mergers (e.\,g.\, \citealt{burns08} and references therein).
More generally speaking, the question arose whether the observed distribution of clusters was due to a primordial division into the two classes or rather to evolutionary differences during the history of the clusters.  \\
For instance \citet{mccarthy04, mccarthy08} suggested that early episodes of non-gravitational pre-heating in the redshift range $1<z<2$ may have increased the entropy of the ICM of some proto-clusters which did not have time to develop a full cool core. \citet{burns08} suggested that while mergers cannot destroy simulated cool cores in the local Universe, early major mergers could have destroyed nascent cool cores in an earlier phase of their formation $(z<0.5)$. \\
However, the ``evolutionary'' scenario, where recent and on-going mergers are responsible of the CC-NCC dichotomy, has been continuously supported by observations. Indeed, correlations have been shown between the lack of a cool core and several multi-wavelength indicators of on-going dynamical activity (e.\,g.\, \citealt{sanderson06,sanderson09b} and \citealt{leccardi10}). \\
Giant radio halos are the most spectacular evidence of non thermal emission in galaxy clusters (\citealt{ferrari08} for a recent review). Over the last years, there has been increasing collective evidence in the literature that they are found in clusters with a strong on-going dynamical activity (e.\,g.\,\citealt{buote01} and \citealt{govoni04}) suggesting that mergers could provide the energy necessary to accelerate (or re-accelerate) electrons to radio-emitting energies \citep{sarazin02,brunetti09}. Recently, the connection between radio halos and mergers has been quantitatively confirmed on a well-defined statistical sample by \citet{cassano10}. \\
In the framework of ``evolutionary'' scenarios, mergers are also responsible of the CC-NCC dichotomy. Therefore, we expect mergers to cause a relation between the absence of a cool core and the presence of a giant radio halo. The aim of this work is to assess statistically the presence of this relation and to test our interpretation of the origin of the CC-NCC distribution.

\section{The sample and data preparation}
The choice of the sample is an important part of this project because we do not want to introduce ``selection effects'' which could alter the distribution between the absence of a cool core and the presence of a radio halo. We started from the ``GMRT radio halo survey'' \citep{venturi07, venturi08}: a deep pointed radio survey of clusters selected from X-ray flux limited \emph{ROSAT} surveys (REFLEX and eBCS), with $z=0.2-0.4$, $L_X>5\times10^{44}\, \rm{ergs}\,\rm{s}^{-1}$ and $-30\degr<\delta<60\degr$. For the clusters of this sample, \citet{venturi08} could either detect extended radio emission or put strong upper limits on it.
We then looked in the \chandra and \xmmn archives for observations of the clusters in the GMRT RH sample, excluding the three objects with mini radio halos. We preferentially used \chandra observations in order to exploit the better angular resolution but we discarded observations with less than 1500 counts in each of the regions from which we extract spectra (see Sec.\,\ref{CC-est}), moving to \xmmn when available. Our final sample consists of 22 clusters with available X-ray observations. Of these clusters, 10 are ``radio-loud'' (hosting a giant radio halo obeying the well known relation between the radio power at $1.4$ GHz, $P_{1.4}$, and the X--ray luminosity $L_X$) and the remaining 12 are ``radio-quiet'',  showing no indication of extended central radio emission and well separated in the $P_{1.4}-L_X$ plane (see \citealt{brunetti09} for a detailed discussion of this distribution). We note here that our sample of ``radio--loud'' clusters is composed of all the clusters with a confirmed giant radio halo in \citet{venturi08}, with the addition of A697 and A1758 which were classified as ``candidate halos'' and were confirmed later \citep{macario10,giovannini09}. \\
\chandra and \xmmn observations are reduced using our standard procedures  \citep{gasta09,rossetti10} with the analysis packages CIAO $4.1$ (with \emph{Caldb} $4.1.1$) and SAS $9.0$ respectively. More details on the analysis will be provided in thep aper which will be soon submitted.

\section{Cool core estimators}
\label{CC-est}
For each of the clusters in our sample we have calculated two estimators of the core state: the central entropy $K_0$ \citep{cava09} and the pseudo-entropy ratio $\sigma$ \citep{leccardi10}. $K_0$ is derived from the fit of the entropy profile with the model $K_0+K_{100}(r/100\,\rm{kpc})^{\alpha}$.
When available, we have used the values reported in the ACCEPT catalogue\footnote{http://www.pa.msu.edu/astro/MC2/accept/}. For the 4 objects whose \chandra observations were not public at the time of the compilation of ACCEPT, we have extracted the entropy profile following the same procedure as \citet{cava09} and fitted it to recover $K_0$. \\ 
The pseudo-entropy ratio is defined as $\sigma=(T_{IN}/T_{OUT})*(EM_{IN}/EM_{OUT})^{-1/3}$, where $T$ is the temperature, $EM$ is the emission measure (XSPEC normalization of the mekal model divided by the area of the region). The $IN$ and $OUT$ regions are defined with fixed fraction of $R_{180}$ ($R<0.05R_{180}$ for the $IN$ region and $0.05R_{180}<R<0.2R_{180}$ for the $OUT$ region).
To measure $\sigma$, we applied the procedure in \citet{leccardi10} to \chandra and \xmmn data. 

\section{Results}
\begin{figure}[]
\subfigure{
\hspace{-0.8 cm}
\resizebox{1.1\hsize}{!}{\includegraphics[clip=true]{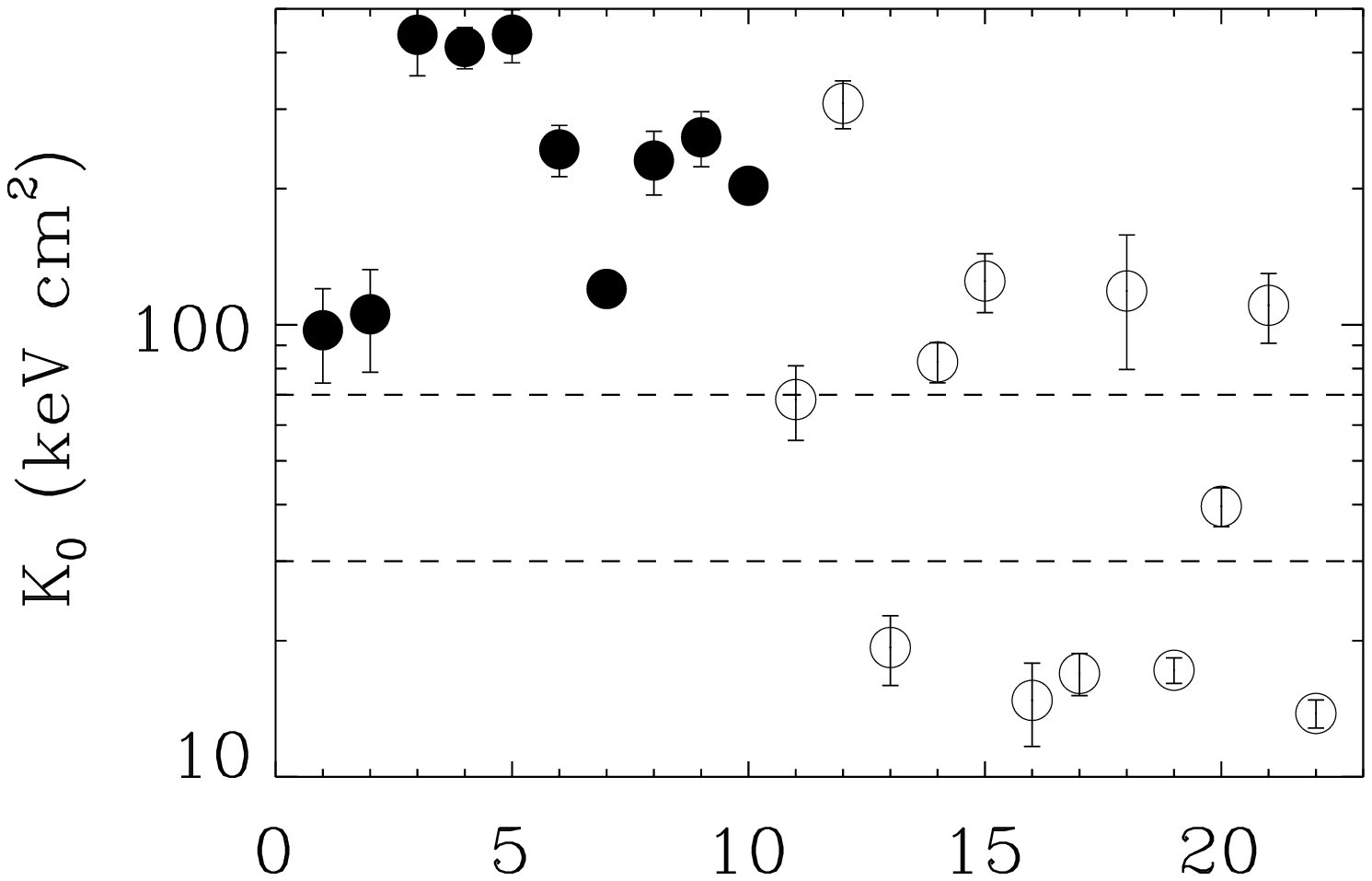}}
}\\
\vspace{-0.8cm}
\subfigure{
\hspace{-0.8 cm}
\resizebox{1.1\hsize}{!}{\includegraphics[clip=true]{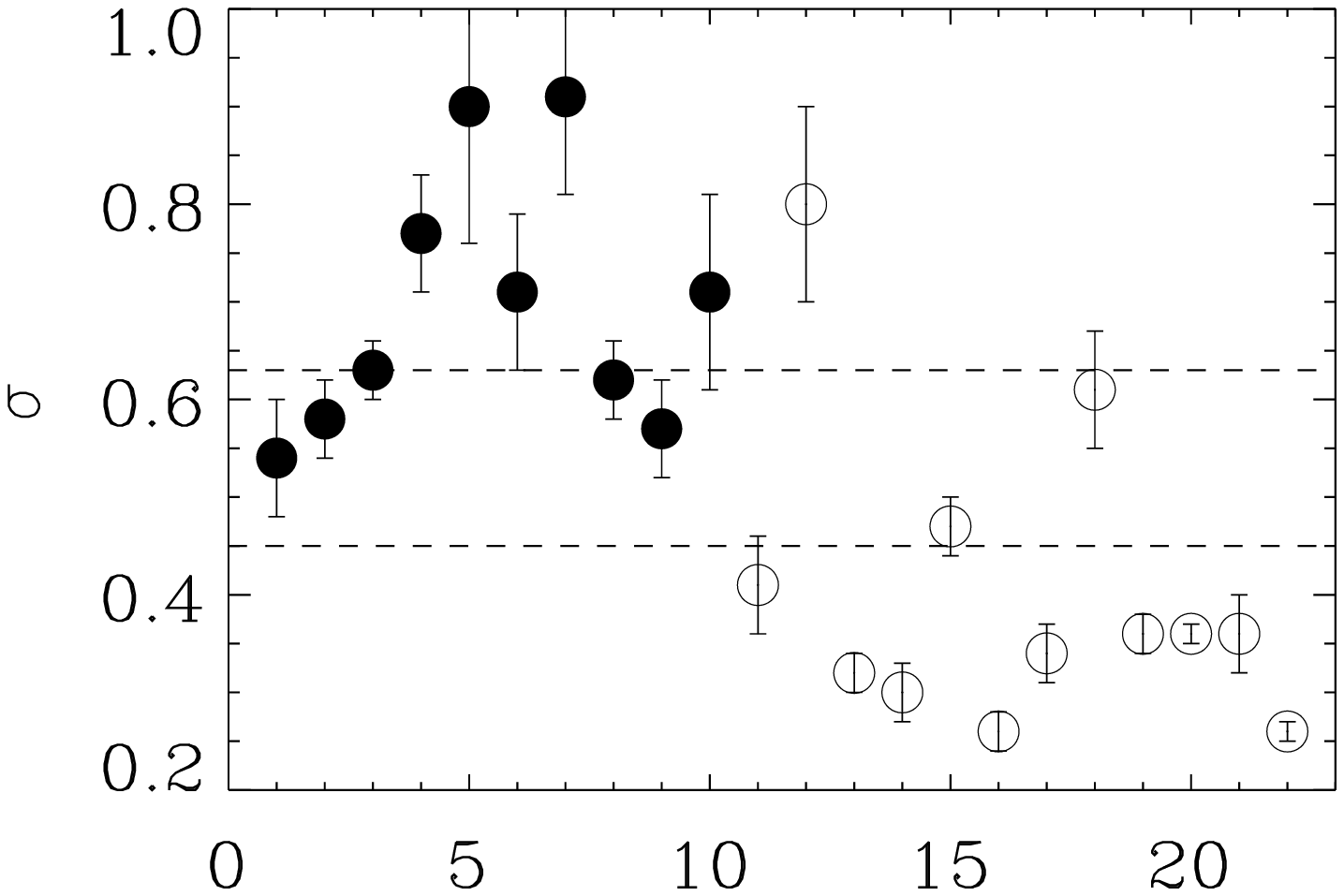}}
}
\caption{Cool core indicators ($K_0$ and $\sigma$) for all the clusters in the sample. Filled symbols are ``radio-loud'' clusters while open symbols are radio-quiet.}
\label{fig_ind}
\end{figure}

\citet{cava09} have shown that the central entropy $K_0$ is a good indicator of the core state. On the basis of Fig.\,6 in their paper, we divided the clusters population into three classes: CC ($K_0<30\,\rm{keV}\,\rm{cm}^2$), NCC ($K_0>70\,\rm{keV}\,\rm{cm}^2$) and intermediate objects (INT $30<K_0<70\,\rm{keV}\,\rm{cm}^2$) where the tails of the two distribution overlap. Using this classification, we found that all ``radio-loud'' clusters are classified as NCC while ``radio quiet'' objects belong to all three classes\footnote{A qualitatively similar result has reported also by \citet{ensslin10}.} (Fig.\,\ref{fig_ind} upper panel). Because of the relatively low number of objects in our sample, we have to verify our result with Monte Carlo simulations to exclude that it comes out just from statistical fluctuations. Therefore we have calculated the mean $K_0$ of our sample of radio loud clusters ($K_0=254 \pm 13\,\rm{keV}\,\rm{cm}^2$) and compared it with the distribution of the mean $K_0$ of 10 clusters randomly selected in the ACCEPT archive. We found that the probability of finding by chance a mean $K_0$ larger than the value of the radio-loud sample is only $0.009\%$ ($0.005\%-0.027\%$ at 1 $\sigma$).\\
We have performed the same analysis using the pseudo-entropy ratios $\sigma$, using the thresholds  in \citet{leccardi10} to divide objects into classes (CC if $\sigma<0.45$, NCC if $\sigma>0.63$ and INT in between). We found that none of the radio--loud clusters is classified as a CC while radio--quiet objects belong to all three classes (Fig.\,\ref{fig_ind} lower panel). As for $K_0$, we have performed a Monte Carlo simulation, calculating the mean of our ``radio-loud'' sample ($\sigma=0.69 \pm 0.02$) and comparing it with the distribution of the mean of 10 randomly selected values in the sample of \citet{leccardi10}. We found a chance probability of finding a mean value larger than the observed value of $0.26\%$ ($0.02\%-1.96\%)$ if we consider the $68\%$ errors on the mean $\sigma$).\\
If we plot our results in the $K_0$ vs $\sigma$ plane, there is a quadrant of the plane (defined as $K_0<74\,\rm{keV}\,\rm{cm}^2$ and $\sigma<0.49$) where no radio-halo cluster is found. We performed a Monte Carlo simulation randomly picking out 10 clusters in the total sample and found that only in 15 out of $10^5$ trials no cluster is found in the selected quadrant ($p=0.015\%$ of being a statistical fluctuation).\\

\section{Conclusion}
We found robust statistical indications for a relation between the absence of a cool core (as indicated by both $K_0$ and $\sigma$) and the presence of a giant radio--halo. Despite the relatively low number of objects in our sample this result is statistically significant, as shown by our Monte Carlo simulations from which we computed the probability of a chance result to be lower than $2\%$ (even in the worst case). Moreover these results have been obtained with a well defined sample, without selection biases towards NCC clusters: the ``radio-loud'' objects we have analyzed are all the clusters in the survey with a confirmed radio halo. \\
This result is naturally addressed in ``evolutionary'' scenarios of the CC-NCC dichotomy where recent and on-going mergers are responsible for the disruption of the cool cores and also of powering the radio-emitting population. Conversely, alternative ``primordial'' scenarios would have to explain why radio-halos are found only in NCC object. 

\begin{acknowledgements}
 We wish to thank Chiara Ferrari and the organizers of the NTGC 2010 conference. We acknowledge useful discussion with R.Cassano, G.Brunetti and T.Venturi. We thank P.Humphrey for the use of his \chandra code and K.Cavagnolo for the ACCEPT catalogue.

\end{acknowledgements}

\bibliographystyle{aa}

\end{document}